\documentclass[aps,prl,twocolumn,amsmath,amssymb,superscriptaddress,longbibliography]{revtex4-2}
\usepackage[breaklinks,colorlinks,bookmarks=false,citecolor=blue,linkcolor=red,urlcolor=blue]{hyperref}
\usepackage{blindtext,enumitem,graphicx,dcolumn,color,xcolor,amsmath,braket,bm,changes,chemformula,cleveref,times}
\usepackage{tikz-feynman}

\graphicspath{{figures_prl/}}
\usepackage{blindtext}
\usepackage{changes}
\crefname{equation}{Eq.}{Eqs.}
\Crefname{equation}{Eq.}{Eqs.}
\crefname{figure}{Fig.}{Figs.}
\Crefname{figure}{Fig.}{Figs.}
\crefname{section}{Sec.}{Secs.}
\Crefname{section}{Sec.}{Secs.}

\newcommand{\JD}{J_D}

\begin{document}
%\title{Anomalous thermal broadening in the dimer phase of the Shastry-Sutherland model}
\title{Anomalous thermal broadening in the Shastry-Sutherland model and SrCu$_2($BO$_3)_2$}

\author{Zhenjiu Wang}
\email{zhenjiu.wang@physik.uni-muenchen.de}
\affiliation{Arnold Sommerfeld Center for Theoretical Physics,
University of Munich, Theresienstr. 37, 80333 Munich, Germany}
\affiliation{Max Planck Institute for the Physics of Complex Systems, N\"othnitzer Strasse 38, Dresden 01187, Germany}

\author{Paul McClarty}
\thanks{Z. Wang and P. McClarty contributed equally to the manuscript.}
\affiliation{Max Planck Institute for the Physics of Complex Systems, N\"othnitzer Strasse 38, Dresden 01187, Germany}
\affiliation{Laboratoire Léon Brillouin, CEA, CNRS, Université Paris-Saclay, CEA Saclay, 91191 Gif-sur-Yvette, France}

\author{Dobromila Dankova}
\affiliation{Max Planck Institute for the Physics of Complex Systems, N\"othnitzer Strasse 38, Dresden 01187, Germany}

\author{Andreas Honecker}
%\email{Andreas.Honecker@cyu.fr}
\affiliation{Laboratoire de Physique Th\'eorique et Mod\'elisation, CNRS UMR 8089, CY Cergy Paris Universit\'e, 95302 Cergy-Pontoise, France}%%

\author{Alexander Wietek}
\email{awietek@pks.mpg.de}
\affiliation{Max Planck Institute for the Physics of Complex Systems, N\"othnitzer Strasse 38, Dresden 01187, Germany}

\begin{abstract}
The quantum magnet SrCu$_2($BO$_3)_2$ and its remarkably accurate theoretical description, the spin-$1/2$ Shastry-Sutherland model, host a variety of intriguing phenomena such as a dimer ground state with a nearly flat band of triplon excitations, a series of magnetization plateaux, and a possible pressure-induced deconfined quantum critical point. One open puzzle originating from inelastic neutron scattering and Raman experiments is the anomalous broadening of the triplon modes at relatively low temperatures compared to the triplon gap $\Delta$. We demonstrate that the experimentally observed broadening is captured by the Shastry-Sutherland model. To this end, we develop a numerical simulation method based on matrix-product states to simulate dynamical spectral functions at nonzero temperatures accurately. Perturbative calculations identify the origin of this phenomenon as a small energy scale compared to $\Delta$ between single triplon and bound triplon states at the experimentally relevant model parameters.
\end{abstract}

\date{\today}

\maketitle

\paragraph{Introduction}  One of the most exquisite examples of geometric frustration in quantum magnetism is the spin-$1/2$ Shastry-Sutherland model in which frustrated antiferromagnetic triangular units support an exact dimer covering in two dimensions over a significant range of exchange couplings \cite{ShastrySutherland1981}. Simple though the ground state in this model exemplifies nicely the effect of destructive interference on triangular units that is the heart of geometrical frustration the exploration of which has formed an entire field of research. Today studies of highly frustrated magnets range over rich and complex physics on kagome, pyrochlore, and other lattices \cite{lacroix2011introduction}. There are however further pillars to the fame of the Shastry-Sutherland model. One is the discovery, many years after the original theory paper, of a material SrCu$_2($BO$_3)_2$, that almost perfectly realizes the dimer phase of the model \cite{Kageyama1999,Kageyama1999JPSJ}. This material, albeit one of thousands of magnetic materials, has consistently stood out for the surprises and puzzles that it has generated and continues to produce over twenty years after the first experiments \cite{Kageyama1999,onizuka2000,kodama2002,miyahara2003,takigawa2004,Kodama_2005,Levy2008,sebastian2008,jaime2012,takigawa2013,matsuda2013,haravifard2016,Nomura2023,Kageyama1999JPSJ,nojiri1999,Lemmens2000,kageyama2000,Room2000,gaulin2004,McClarty_2017, wulferding2021,miyahara1999,miyahara69thermodynamic, knetter2000,cepas2001,romhanyi2011,Romh_nyi_2015,Totsuka2001,Knetter2004,Waki_2007,corboz2014,Zayed_2017,Haravifard_2012,Sakurai_2018,Boos2019,Mila2020,Jim_nez_2021,corboz2023,yang2022,lee2019,wietek2019}.
%\cite{Kageyama1999,Kageyama1999JPSJ,nojiri1999, miyahara1999,miyahara69thermodynamic,kageyama2000, knetter2000,onizuka2000,Room2000,cepas2001,Totsuka2001,kodama2002,miyahara2003,Knetter2004,takigawa2004,gaulin2004,Kodama_2005,Waki_2007,Levy2008,sebastian2008,romhanyi2011,Haravifard_2012,jaime2012,takigawa2013,matsuda2013,corboz2014,Romh_nyi_2015,McClarty_2017,Zayed_2017,haravifard2016,Sakurai_2018,Boos2019,lee2019,wietek2019,Mila2020,wulferding2021,Jim_nez_2021,yang2022,corboz2023,Nomura2023}. 
These include the famous series of magnetization plateaux reaching up to around $100$ T \cite{Kageyama1999,onizuka2000,kodama2002,miyahara2003,takigawa2004,Kodama_2005,Levy2008,sebastian2008,jaime2012,takigawa2013,matsuda2013,haravifard2016,Nomura2023}, observations of nearly flat triplon excitations about the dimer phase \cite{kageyama2000, gaulin2004},  IR, Raman and neutron studies exploring the triplons and the tower of bound state excitations \cite{nojiri1999,Lemmens2000,Room2000,kageyama2000,gaulin2004,McClarty_2017,wulferding2021}, topological triplons coming from small exchange anisotropies \cite{cepas2001,romhanyi2011, Romh_nyi_2015,McClarty_2017}, experiments observing a plaquette phase in the material at high pressures and investigations of the nature of the phase transition to this phase \cite{Waki_2007,Haravifard_2012,Zayed_2017,Sakurai_2018,Boos2019,Mila2020,Jim_nez_2021,corboz2023} including the tantalizing possibility of realizing deconfined criticality on the boundary between N\'eel order and the plaquette phase \cite{lee2019,yang2022}. The Shastry-Sutherland model frames all these experimental discoveries and both model and material have provided an important proving ground for new numerical and analytical tools. In fact, studies of SrCu$_2($BO$_3)_2$ have turned out to be almost a microcosm for the development of quantum magnetism as a whole. For example, the low entanglement dimer phase was an attractive target for tensor network methods which now have captured much of the complexity of the magnetic field-induced phase diagram as crystals of condensed bound states \cite{corboz2014}. Related techniques have captured also the thermodynamics of the model \cite{wietek2019}. 
% Series expansion, perturbation theory, ED

In this paper, we consider a further curiosity of SrCu$_2($BO$_3)_2$ $-$ a dramatic broadening of the triplon modes with increasing temperature uncovered by inelastic neutron scattering \cite{kageyama2000,gaulin2004,zayed2014} and corroborated by Raman scattering \cite{Lemmens2000, wulferding2021}. Thermal broadening of excitations is entirely to be expected in any correlated magnetic system. The peculiarity of SrCu$_2($BO$_3)_2$ is its sensitivity to thermal excitations as well as the extent of the effect in the crossover to the paramagnetic state. To be concrete, the triplon excitations are gapped at about $3$ meV or about $30$ K, and significant broadening is observed by around $5$ K, while the Bose factor at this temperature is about $3\times 10^{-3}$ so the triplons are undoubtedly very dilute.  By $15$ K the neutron intensity is broadened into a nearly featureless continuum. In the following, we show how to capture this effect quantitatively and provide insight into the microscopic mechanism that underlies it. On small cluster sizes up to $N=20$ sites, a finite-temperature Lanczos study~\cite{Shawish2005} reported anomalous thermal broadening but left results on larger system sizes desirable. In some sense, this work reflects the natural course of development in understanding low-dimensional magnetic systems using state-of-the-art numerical tools. Tensor network tools were first brought to bear on the ground states \cite{Corboz2013,corboz2014,Nomura2023,Czarnik2021,Vlaar2023} then the thermodynamics \cite{wietek2019,Wietek2021,Wietek2021b,Feng2022,Feng2023} and now it has become feasible to consider the nonzero temperature dynamics of extended two-dimensional frustrated quantum magnets. 

\begin{figure}[t]
\centering
\includegraphics[width=\columnwidth]{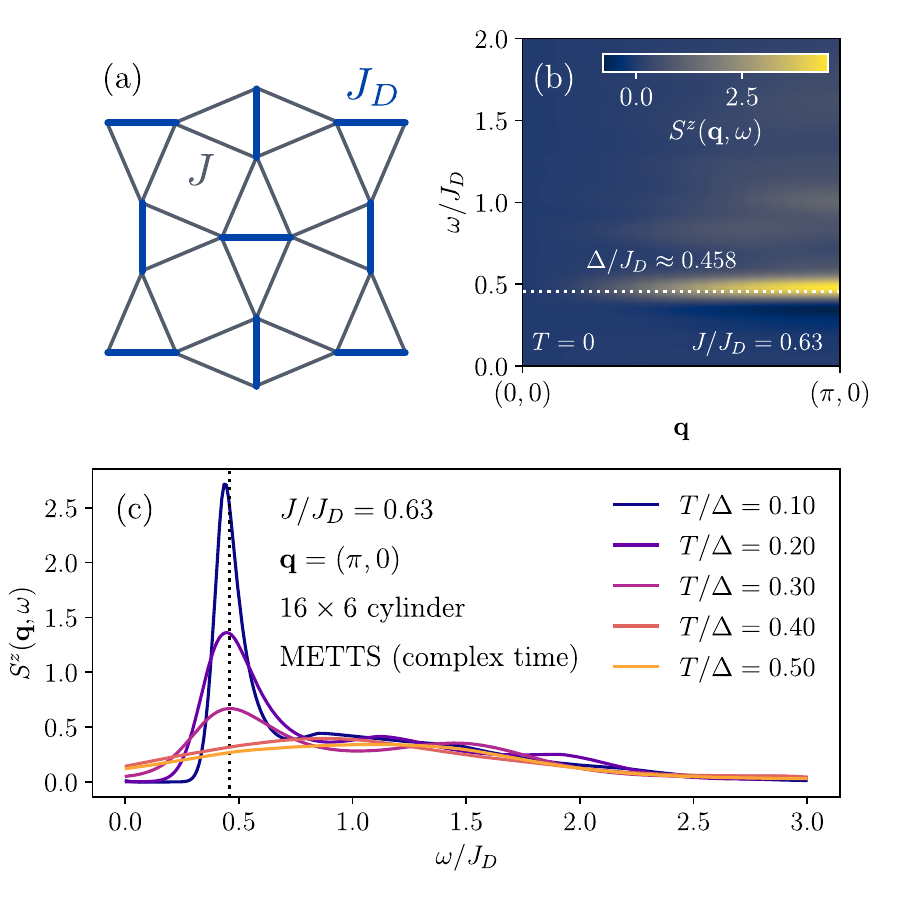}
\caption{(a) Geometry of the Shastry-Sutherland lattice. The intradimer couplings ${\JD}$ are shown in blue and the interdimer couplings $J$ in grey. (b)
Dynamical spin structure factor $S(\bm{q}, \omega)$ at $T=0$ from dynamical DMRG simulations on a $16 \times 6$ cylinder for a cut through the Brillouin zone from $(0, 0)$
to $(\pi, 0)$ exhibiting a flat triplon band at energies around the triplet gap $\Delta \approx 0.458 \JD$.
(c) $S((\pi,0), \omega)$ for various temperatures from the proposed dynamical METTS simulations on a $16 \times 6$ cylinder, employing the analytical complex time algorithm as proposed in the companion paper~\cite{Wang2024b}. The triplon peak at $\Delta$ melts into a broad continuum at temperatures that are a fraction of the triplon gap $\Delta$.
}
\label{fig:ssmodel}
\end{figure}

\paragraph{The Shastry-Sutherland Model and its Experimental Realization}  The Shastry-Sutherland model is a localized spin-$1/2$ model formulated on the lattice illustrated in Fig.~\ref{fig:ssmodel}(a) \cite{ShastrySutherland1981}. The Hamiltonian is given by,
\begin{equation}
H = {\JD}\sum_{\langle i,j \rangle_D} \bm{S}_i\cdot \bm{S}_j + J\sum_{\langle\langle i,j \rangle\rangle} \bm{S}_i\cdot \bm{S}_j,
\end{equation}
where ${\JD}$ denote the intradimer couplings and $J$ the interdimer couplings. When $J=0$ the ground state has isolated singlets living on the ${\JD}$ bonds and canonical excitations are spin $S=1$ triplets in the local dimers. When $J$ is switched on each singlet couples via its two component spins to a single site on neighboring singlets. This geometrical frustration endows the singlets with stability such that the dimer covering survives as an exact eigenstate for all couplings and in the ground state up a $J/{\JD}\approx 0.67$ as determined numerically. 

The lowest-lying excited states, for small $J/{\JD}$, are coupled triplet excitations called {\it triplons}. As the model and ground state are
spin-rotationally symmetric, the triplons are three-fold degenerate. One might expect the coupled triplets to acquire some dispersion on the scale of $J$. However, it turns out that triplon hopping is suppressed by the magnetic frustration with the leading order contribution appearing to $O((J/{\JD})^6)$ \cite{miyahara1999,Zheng1999,knetter2000,knetterJPCM2000,miyahara2003} $-$ so the triplons are very nearly localized even close to the phase boundary out of the dimer phase resulting in a flat band of triplon excitations, cf. Fig.~\ref{fig:ssmodel}(b). 

The material SrCu$_2($BO$_3)_2$ with spin one-half copper ions realizes the Shastry-Sutherland model to a good approximation. At ambient pressure $J/{\JD}$ has been experimentally estimated to be $0.63$ \cite{miyahara69thermodynamic,miyahara2003,wietek2019} so it lies at the edge of the dimer phase.
%{\color{red} *} Indeed, the material exhibits a smooth crossover upon cooling into a non-magnetic state and the triplon modes have been measured using inelastic neutron scattering {\color{red} Ref(s).}.
%Proximity to the phase boundary is corroborated by the presence of low-lying bound states detected using Raman and neutron spectroscopies
%{\color{red} Ref(s).}.
%{\color{red} [AH: the preceding sentences (since *) are redundant with other parts of the manuscript and could thus be shortened, if we are shory of space.]}
Observations of the triplons reveal the bandwidth to be about one-tenth of the gap and therefore significantly larger than in the pure Shastry-Sutherland model. This indicates the presence of small anisotropies that are known to be predominantly Dzyaloshinskii-Moriya or antisymmetric exchange that we neglect here. Our results imply that these small corrections to the pristine Shastry-Sutherland model play a negligible role in the thermal broadening to which we now turn. 

The sensitivity of the quantum states of SrCu$_2($BO$_3)_2$ to finite temperatures was first observed from the washing out of magnetization plateaux at around 1~K \cite{Kageyama1999}. Later the neutron scattering intensity of the single triplon modes was seen to fall off faster with increasing temperature than would be expected on the basis of the $35$~K gap \cite{kageyama2000,gaulin2004,zayed2014}. Indeed, the triplons are almost completely washed out by about $10$ K leaving a broad continuum of intensity. This behavior is consistent with Raman measurements that are more sensitive to singlet intensity \cite{Lemmens2000,wulferding2021}. It has been suspected for a long time that the unusual temperature dependence originates from the delicate nature of the frustration-induced dimer formation and, in particular, that only a very dilute concentration of thermal triplet states is sufficient to delocalize the dimers. 

\begin{figure*}
    \centering
    \includegraphics[width=\textwidth]{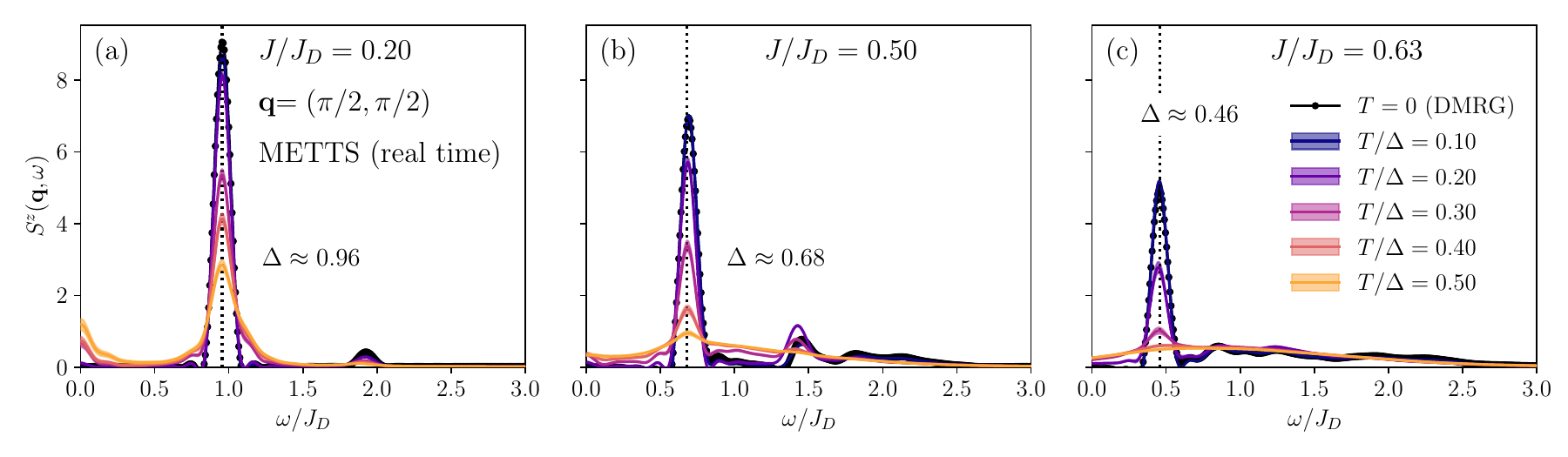}
    \caption{Dynamical spin spectral functions evaluated at momentum $\bm{q} = (\pi / 2, \pi / 2)$ from dynamical METTS simulations on the $16\times4$ cylinder.  Results at temperatures which are a fixed ratio of the triplet gap $\Delta$ as in \cref{eq:gap} are shown. The triplet gap $\Delta$ is shown as the dashed line. (a) $J/{\JD}=0.2$ (b) $J/{\JD}=0.5$ (c) $J/{\JD}=0.63$. Whereas the dominant peak close to the triplet gap is only weakly broadened for $J/{\JD}=0.2$ it completely disappears at small fractions of the triplet gap for $J/{\JD}=0.63$ in (c).    }
    \label{fig:spectra_tdep_3panel}
\end{figure*}

\paragraph{Thermal broadening from dynamical METTS}
We will now demonstrate the dynamics of the pure Shastry-Sutherland model to accurately capture the effect of the triplon thermal broadening in SrCu$_2($BO$_3)_2$ with close agreement with experimental measurements. Our simulation is based on a numerical technique for evaluating dynamical spectral functions at nonzero temperature, based on the idea of minimally entangled typical thermal states \cite{White2009,Stoudenmire2010,Wietek2021,Wietek2021b}. In the companion paper \cite{Wang2024b}, we introduce this technique in detail and benchmark it against more traditional techniques where such a comparison is possible. The principal advantage of dynamical METTS is that it allows one to address significantly larger system sizes than previously possible. The method is performed in two distinct modes. With the real-time evolution algorithm, we directly simulate the dynamical correlation function,
\begin{equation}
\mathcal{C}_{AB}(t) = \langle A(t)B \rangle_\beta = \langle e^{iHt}A e^{-iHt} B\rangle_\beta,
\end{equation}
at nonzero temperature ($T=1/\beta$) up to a final time $\Omega$, after which a Fourier transform yields the desired spectral function. The dynamical correlation functions can be fully converged on $W=4$ cylinders at all investigated temperatures for the Shastry-Sutherland model. Here, we chose a simulation time horizon of $\Omega/{\JD}=50$.

This method yields highly accurate results for smaller cylinders, where even the temperature dependence of secondary and tertiary peaks can be resolved. However, since there are limitations in system size we introduce a complex-time evolution algorithm, where the dynamical correlation function is simulated on a contour in complex time coordinates and the spectral function is obtained via stochastic analytic continuation \cite{Silver_1990,Sandvik_SAC_98,beach2004identifying,SHAO20231}. The fact that the correlation function is not just simulated on the imaginary-time axis yields improvements in the ill-posedness of the analytical continuation. For all the necessary details, we refer to the companion paper \cite{Wang2024b}.

Figure~\ref{fig:spectra_tdep_3panel} shows the dynamical structure factor $S(\bm{q},\omega)$ calculated for fixed $\bm{q}=(\pi/2,\pi/2)$ and for different temperatures and various $J/{\JD}$. For $J/{\JD} = 0.2$, the principal peak at $\omega/\JD=0.96$ corresponds to the ground state to single triplon transition and there is a secondary peak at twice this energy coming mainly from the free two-triplon states. As the temperature increases the amplitude of both peaks decreases and both broaden and, at the same time, quasi-elastic intensity appears. 

For larger values of $J/{\JD}$ the single triplon peak comes down in energy and for a fixed temperature it is broader for larger values of $J/{\JD}$. Meanwhile, the two-triplon sector broadens into a continuum extending to both higher and lower energies. For $J/{\JD}=0.63$ corresponding to the material the dynamical structure factor is a featureless continuum above around $T/\Delta=0.4$ corresponding to about $14$ K.

\begin{figure}[tb!]
    \centering
    \includegraphics[width=\columnwidth]{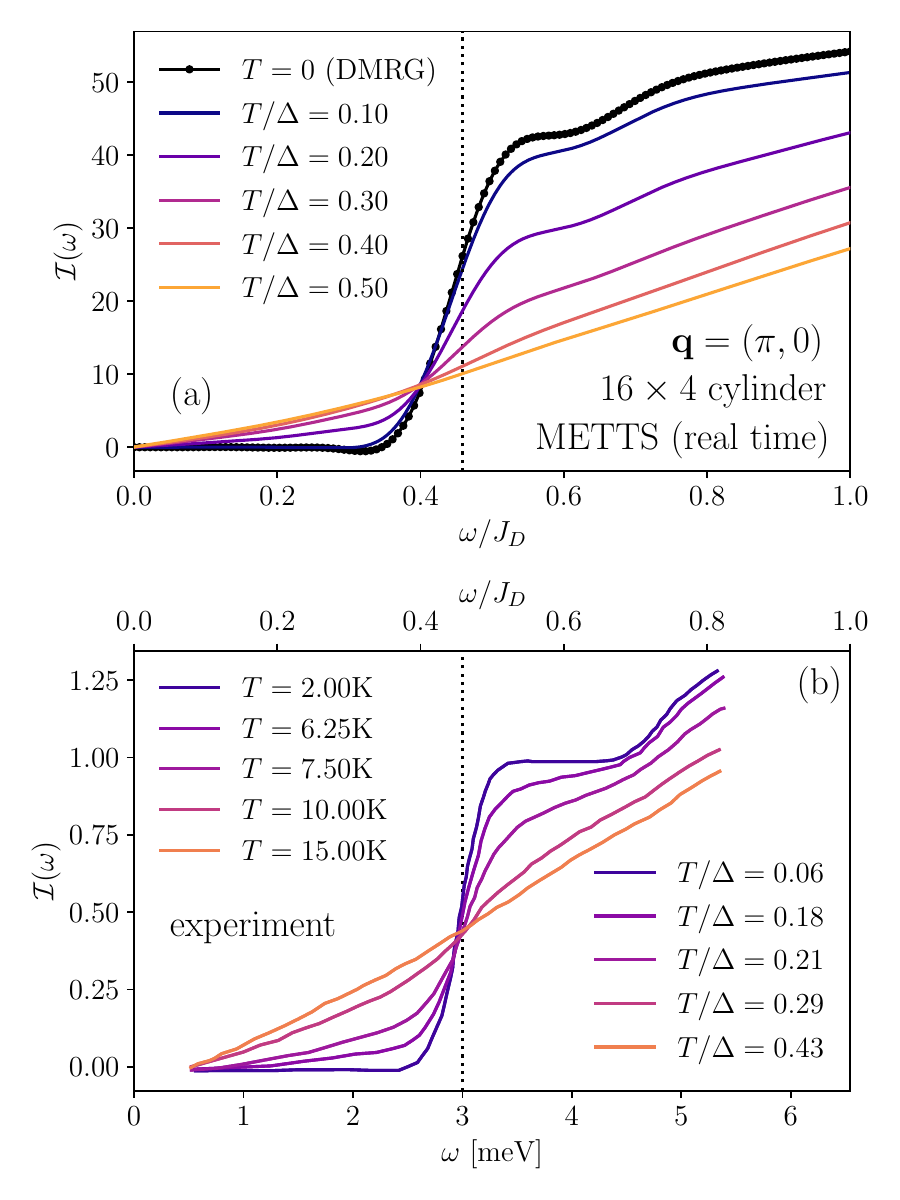}
    \caption{Temperature dependence of the cumulative spectral weight as a function of energy. The lower panel is inelastic neutron scattering data (taken from Ref.~\cite{zayed2014}) on a powder sample. The data is momentum-integrated and resolved in energy. The data at five different temperatures reveals the progressive broadening of the central peak. The top panel is our numerical result for the cumulative spectral weight at
    momentum $(\pi,0)$. }
    \label{fig:comparison}
\end{figure}

On the $16\times 6$ cylinder we analogously observe the melting of the main triplon peak at temperatures $T/\Delta=0.4$ in \cref{fig:ssmodel}(c). There, results have been obtained using the complex-time evolution algorithm explained in the companion paper \cite{Wang2024b}.

%\section{Finite temperature dynamics using METTS}

We now make a more direct comparison of the numerical results and the experiment. Figure~\ref{fig:comparison} shows the cumulative spectral weight up to energy $\omega$ for different temperatures. The top panel is the numerical data at fixed momentum and the lower panel is the experimental data taken from Ref.~\cite{zayed2014}. The lower temperature data shows the single triplon peak as a rapid upturn in both numerics and experiment. At temperatures of about $0.4$ of the triplon gap, the cumulative spectral weight increases almost linearly corresponding to an almost featureless continuum. This plot demonstrates that the numerical technique captures the thermal broadening in the material. In particular, the degree of broadening at a given temperature scale coincides between
the simulation and the experiment. Ref.~\cite{zayed2014} points out that the single triplon peak appears to have one sharp component whose amplitude decreases with temperature and a broad temperature-dependent component, which is consistent with our numerics.

A few comments and caveats are necessary at this point. One is that the zero temperature peak is a delta function in principle. This is not realized in the numerics because of the inevitable finite time cutoff in the dynamics. The experimental single triplon peak also has a width at very low temperatures owing to instrumental resolution. Secondly, one plausible lesson to be drawn from the excellent agreement between theory and experiment is that the relevant physics is rather local. As we shall argue this originates from the almost perfect localization of the single triplon modes combined with the fact that small system sizes already capture the broad spectrum of bound-state modes as the relevant states have a short length scale. Finally, one might be concerned that the material has couplings beyond the Heisenberg model and that these may contribute to the thermal broadening. 
%% I don't think we need that:
%At first sight, this may sound reasonable as the frustration-induced single triplon localization is broken on the scale of these perturbations.
The good agreement between the numerical results and experiment is nevertheless suggestive that the pure Shastry-Sutherland model is largely responsible for the physics. We argue below that indeed the relevant scales are those coming from the Heisenberg model. 

\begin{figure}[t]
    \centering
    \includegraphics[width=\columnwidth]{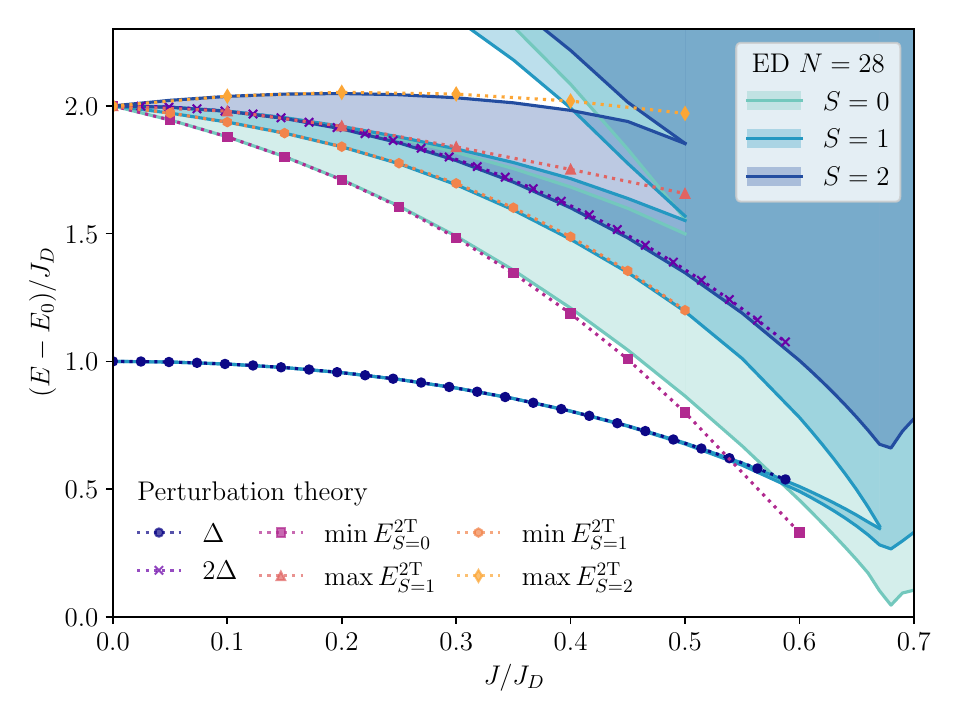}
    \caption{
    Comparison of energy gaps between perturbation theory (PT) to third order in $J/J_{\rm D}$ and exact diagonalization (ED). The shaded regions show continua of spin-$S$ states on a $N=28$ cluster ($S=0$ and $1$ data adapted from Ref.~\cite{wietek2019}). The triplet gap $\Delta$ and the gap to the lowest $S=0$ excitations agree well between PT and ED. At $J/{\JD} \gtrsim 0.6$ the lowest excited states are bound states of two triplons with $S=0$, invisible in the spin structure factor but thermally activated prior to the triplon excitations for the experimentally relevant parameters $J/{\JD}=0.63$.}
    %{\color{red} [AH: the ED discussion needs to be revised. Shaded regions are boundaries of continua, as determined from $N=28$ ED. However, some of the lines are just, e.g., multiples of the $S=1$ excitation, and {\em not} numerical results for $S>1$.]}
    \label{fig:ed_pt}
\end{figure}

{\it Physics of thermal broadening} $-$ Having seen that the Shastry-Sutherland model in a non-perturbative analysis leads to thermal broadening similar to that seen in SrCu$_2($BO$_3)_2$ we now discuss the microscopic origin of this phenomenon. 

To set the scene we briefly review some pertinent features of the model. As noted above, interactions mediated by the exchange $J$ lead to a triplon dispersion, to leading order, only at the sixth order in the coupling \cite{miyahara1999,Totsuka2001,miyahara2003} so, for the parameters corresponding to the material, the triplon modes are expected to be nearly flat. The flatness of the single particle modes further implies that the two-triplon continuum is very narrow in energy. The interactions, however, do have a significant effect on the triplon energy which is renormalized downwards as 
\begin{equation}
\label{eq:gap}
% \Delta = J_{D}\left( 1 - \left( \frac{J}{{\JD}} \right)^2 - \frac{1}{2} \left( \frac{J}{{\JD}} \right)^3 - \frac{1}{8} \left( \frac{J}{{\JD}} \right)^4 \right)
\Delta = J_{D}\left[ 1 - (J/{\JD})^2 - \frac{1}{2}(J/{\JD})^3 - \frac{1}{8}(J/{\JD})^4 \right]
\end{equation}
to $4$th order in perturbation theory \cite{miyahara1999,miyahara2003}. 

Complex collective physics in this model originates from the formation of bound states of triplons. The lowest-lying of these are bound states of two triplons which occur in the $S=0, 1$, and $2$ sectors of which the former contain those of lowest energy. Splitting of the different angular momentum sectors grows like $O(J)$ and the bandwidth in each sector like $O(J^2)$ so the localizing effects of frustration on single triplons are absent in the two-triplon sector. This disparity between the single and two-particle states sets this model apart from typical frustrated magnets. The gap between the single triplons and the lowest bound state excitations closes in the vicinity of $J/{\JD} = 0.6$. Remarkably real-space perturbation theory to third order in $J/{\JD}$ leads to bound states with a bandwidth in good agreement with exact diagonalization. This is illustrated in Fig.~\ref{fig:ed_pt} which reveals that the value of $J/{\JD}$ at which the single and two-particle levels cross is slightly underestimated in the perturbation theory~\cite{supp}. 

\begin{figure}[t]
    \centering
    \includegraphics[width=\columnwidth]{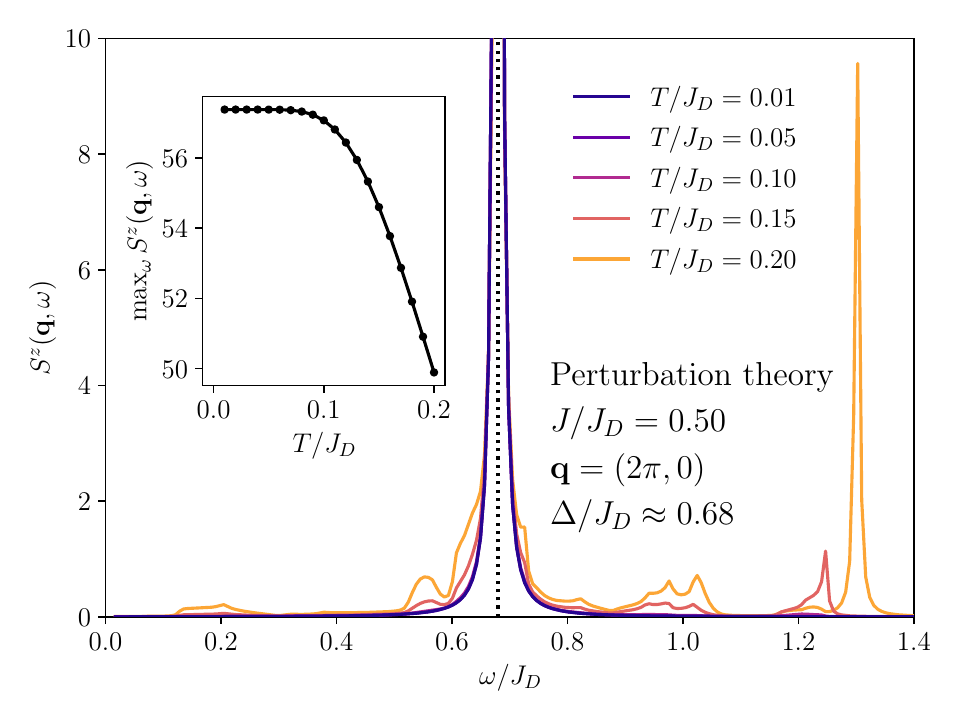}
    \caption{Dynamical structure factor at $J/{\JD}=0.5$ and $\bm{q}=(2\pi,0)$ computed from the low-temperature expansion described in the main text and supp. mat.~\cite{supp}. The inset shows the evolution of the peak height with increasing temperature.}
    \label{fig:lte}
\end{figure}

At zero temperature, the dynamical structure factor has been computed perturbatively in Ref.~\cite{Knetter2004} and has a delta function peak at the single triplon energy at least when there is a separation of energies between the one and two-triplon states.
At finite temperatures, one may formulate the problem as a calculation of self-energy $\Sigma(\bm{q},\omega)$ to obtain susceptibility
\begin{equation}
\chi^{zz}(\bm{q},\omega) = \frac{D(\bm{q},\omega)}{1-D(\bm{q},\omega)\Sigma(\bm{q},\omega)} ,
\label{eq:dyson}
\end{equation}
where $D$ is the single triplon propagator and the $zz$ component of the dynamical structure factor is chosen without losing generality. From this, we get the dynamic structure factor
\begin{equation}
S^{zz}(\bm{q},\omega) = - \frac{1}{\pi} \frac{1}{1-e^{-\beta\omega}} {\rm Im} \left[ \chi^{zz}(\bm{q},\omega) \right].
\end{equation}

We compute the self-energy within a low-temperature expansion by matching the leading order appearance of the self-energy from Eq.~(\ref{eq:dyson}) with those terms from the spectral representation of the dynamical correlator \cite{Essler2008,James2008,supp}
\begin{equation}
\langle S^z_i(\tau) S^z_j(0) \rangle = \frac{1}{Z}\sum_{m,n} e^{-\beta E_m} \langle m \vert S^z_i(\tau) \vert n \rangle \langle n \vert S_j^z(0) \vert m \rangle
\end{equation}
that contribute at low temperatures. This ends up meaning that we compute a re-summed self-energy of the form
\begin{align}
&\Sigma(\bm{q},\omega) \nonumber \\ &\ =   D^{-2} \left( C_{11} + C_{12} + C_{21} - e^{-\beta\epsilon} D - Z_1 (1-e^{-\beta\epsilon}) D \right) 
\end{align}
where $C_{mn}$ are defined through $\chi^{zz}(\bm{q},\omega)= (1/Z)\sum_{mn}C_{mn}$ where the terms in the sum refer to $m/n$-triplon states and $Z_1$ is the single triplon contribution to the partition sum. The contribution $C_{11}$ vanishes when working consistently to third order in $J/J_{\rm D}$ as the single triplons are dispersionless. The $C_{12}$ and $C_{21}$ contributions are split into pieces that come from free two-triplon states and from the bound states. The former contains a part that scales with the number of unit cells, $N$, which cancels with $Z_1 (1-e^{-\beta\epsilon}) D$. The central contribution to the broad response in energy comes from the bound states. 

Fig.~\ref{fig:lte} shows the resulting dynamical structure factor for $J/{\JD}=0.5$ and for $\bm{q}=(2\pi,0)$ where the single triplon intensity is maximal. Several temperatures are plotted between $T/{\JD}=0.01$ up to $0.2$ ($T/\Delta \approx 0.3$). Notably, the delta peak corresponding to the single triplon mode remains but its amplitude decreases with increasing temperature as shown in the inset. In addition to the single triplon peak, there is a broad response originating from the bound states that, as we have mentioned, have a bandwidth of the order of $J$. This broad component to the dynamical structure factor is bounded, for  $J/{\JD}=0.5$, by the gap between the triplon mode and the lowermost and uppermost bound state modes $-$ namely $\Delta/{\JD}=0.12$ and about $1.5$. 

To summarize, the perturbation theory gives an account of the broad inelastic response appearing at energies much smaller than the single triplon gap. Central to this is the ``fine-tuning" in SrCu$_2($BO$_3)_2$ such that it lies close to a phase boundary and the bound states have anomalously low energy. In  ``typical" gapped quantum magnets, two-particle states would arise at around $2\Delta$ resulting in an inelastic response from energy $\Delta$. In contrast, in ``typical" gapless quantum magnets, a continuum of states and broadening are both to be expected at {\it zero} temperature so that effects of finite temperature will tend to be quantitative, not qualitative. In this way, we can understand why thermal broadening SrCu$_2($BO$_3)_2$ stands out among quantum magnets.

\paragraph{Conclusion}
We investigated the origin of the anomalous thermal broadening observed in neutron scattering experiments of SrCu$_2($BO$_3)_2$. By introducing a matrix-product state-based technique based on minimally entangled typical thermal states we demonstrated that this effect is accurately captured by the Shastry-Sutherland model on cylinders up to width $W=6$. Moreover, we provide an intuitive explanation where bound states of two triplons proliferate below the single triplon gap at the experimentally relevant model parameters $J/{\JD}=0.63$. %\textcolor{blue}{In a broader sense, this work contributes to our understanding of anomalous thermal broadening of magnetic excitations that has received attention using an array of powerful tools \cite{}.} 
By demonstrating the feasibility of studying finite-temperature dynamics using tensor network methods, this work paves the way for future investigations of frustrated quantum magnets at non-zero temperatures. 

\begin{acknowledgements}
A.H.\ and A.W.\ are grateful to 
P.\ Corboz, F.\ Mila, B.\ Normand, and S.\ Wessel
for previous related collaborations and discussions. A.W. acknowledges support by the DFG through the Emmy Noether program (Grant No.\ 509755282). 
Z.W.\ was supported by the FP7/ERC Consolidator Grant QSIMCORR, No.\ 771891 and  by the Deutsche Forschungsgemeinschaft (DFG, German Research Foundation) under Germany's Excellence Strategy -- EXC-2111 -- 390814868.  
\end{acknowledgements}
\bibliography{main}
\end{document}